\documentclass[a4paper]{spie}  %>>> use this instead for A4 paper
\usepackage[a4paper,top=2.54cm,bottom=4.94cm,left=1.93cm,right=1.93cm]{geometry} 
\usepackage{amsmath,amsfonts,amssymb}
\usepackage{graphicx}
\usepackage[colorlinks=true, allcolors=blue]{hyperref}
\usepackage{caption}
\usepackage{subcaption}

\title{Radiation effects and noise evolution in NewAthena WFI flight-production sensors}

\author[a]{Valentin Emberger}
\author[a]{Johannes Müller-Seidlitz }
\author[a]{Luisa Ostler}
\author[c]{Wolfgang Treberer-Treberspurg}
\author[a]{Robert Andritschke}
\author[a]{Annika Behrens}
\author[a]{Günter Hauser}
\author[b]{Peter Lechner}
\author[a]{Astrid Mayr}
\author[a]{Leonie Sommer}

\affil[a]{Max-Planck-Institute for Extraterrestrial Physics, Giessenbachstr 1, 85748 Garching, Germany }
\affil[b]{Semiconductor Laboratory of the Max-Planck-Society , Isarauenweg 1, 85748 Garching, Germany}
\affil[c]{Univ. of Applied Sciences Wiener Neustadt, Johannes-Gutenberg-Straße 3, 2700 Wiener Neustadt, Austria}

\authorinfo{Send correspondence to V. Emberger}

\begin{document} 
\maketitle

\begin{abstract}
The Wide Field Imager (WFI), one of the two instruments on ESA’s next large X-ray observatory NewAthena, is designed for imaging spectroscopy in the 0.2–15 keV range, combining a large field of view with high count-rate capability. Its focal plane is equipped with back-illuminated DEPFET (Depleted p-channel field-effect transistor) sensors that offer high radiation tolerance and provide near Fano-limited energy resolution. Achieving this performance requires an exceptionally low readout noise, with about 3 electrons ENC expected at beginning of life. Consequently, the devices are highly sensitive to radiation-induced changes in noise behavior. In this work, we investigate the impact of both total non-ionizing dose (TNID) and total ionizing dose (TID) on the relevant noise components, including their temperature dependence.
A detector module containing a 64$\times$64-pixel sensor from a flight-production wafer was irradiated with 62.4 MeV protons at the MedAustron accelerator facility in Wiener Neustadt to a total dose equivalent to 2.6\,$\cdot$\,10\textsuperscript{9} 10-MeV-protons/cm². The detector was fully biased and operated throughout the irradiation and subsequent measurements, maintaining the nominal operating temperature of 213 K. To study short-term annealing behavior at low temperature, a second, identical module was exposed to a comparable proton dose within a much shorter timescale by exploiting the available high beam flux. TID effects were investigated separately by irradiating another device with 17.4 keV Mo-K\textsubscript{$\alpha$} X-rays to a total dose of 15 Gy.
We report the resulting changes in readout noise, dark current, and threshold voltage, and compare them with results from an earlier irradiation campaign using pre-flight sensors. Implications for the instrument’s required operating temperature and its expected end-of-life performance are discussed.

\end{abstract}


\section{INTRODUCTION}
\label{sec:intro}  % \label{} allows reference to this section

 The focal plane of the WFI\cite{valeria24} instrument for NewAthena\cite{Cruise24} will consist of DEPFET sensors. The current detector design, together with recent results from electrical and performance characterization, is described in Refs.~\citenum{mibo19, mibo22, nom20, joschi24}.

As the mission approaches adoption, the WFI team aims to demonstrate the performance of the instrument subsystems under representative environmental conditions. For the WFI camera head, which is centered around a DEPFET sensor, the dominant degradation mechanism in orbit is expected to be an increase in dark current caused by displacement damage from high-energy protons. To limit the resulting radiation-induced noise contribution to acceptable levels, the DEPFET sensors must be operated at low temperatures. The current baseline foresees an operating temperature of 213~K at the beginning of the mission (Beginning of Life, BOL), with the capability to reduce the temperature to 193~K by the end of the nominal mission (End of Life, EOL). This operating range addresses two competing requirements: maintaining a sufficiently high temperature to minimize contamination of the sensor during the early phase of the mission, while keeping the temperature low enough to suppress thermally generated charge in the large sensitive volume of the device.

The effects of proton irradiation were investigated using two sensors from an NewAthena flight-production wafer. The irradiation campaign is described in Section~\ref{sec:proton}. Previous radiation studies have shown that DEPFET sensors respond to TID not only through shifts in threshold voltage but also through an increase in read noise. However, the physical origin of this additional noise contribution has remained unclear. \

Proton irradiation inevitably introduces both TNID and TID. However, the proton energy spectrum available at irradiation facilities cannot exactly reproduce the spectrum expected in orbit. Consequently, the relative contributions of TNID and TID differ from those encountered in the space environment. Furthermore, the temporal evolution and temperature dependence of the corresponding noise contributions are markedly different. Therefore, a dedicated TID irradiation campaign is required in addition to the proton irradiation study. The TID test, performed with a separate sensor from the flight production, is presented in Section~\ref{sec:TID}. Since this irradiation introduces only ionizing damage, it allows the effects of TID to be isolated from those of TNID. Combining the results of both irradiation campaigns provides a more complete understanding of the sensor response to the space radiation environment and enables predictions of the detector performance throughout the mission lifetime.

A particular focus of both irradiation studies is the identification of the noise components affected by radiation damage and the characterization of their evolution with time and temperature over the relevant operating range. Sections~\ref{sec:results} and~\ref{sec:results2} discuss the post-irradiation evolution of the individual noise components and assess the detector performance expected for the WFI on NewAthena.

\section{THEORY}
\label{sec:theory}  % \label{} allows reference to this section

The WFI instrument is supposed to achieve a very low readout noise of $\sim$3 electrons Equivalent Noise Charge (e\textsuperscript{-}  ENC) across the complete readout chain. There are three main contributors to the total read noise, the detector, the Detector Electronics (DE) and, the Ground Segment (GS):
\begin{equation} \label{eq:tot_noise}
\sigma_{total}^2 = \sigma_{detector}^2 + \sigma_{DE}^2 +\sigma_{GS}^2 
\end{equation}
These can be divided into numerous sub components, estimations for each have been derived from measurements or analysis.  A detailed budget is presented in Ref.~\citenum{astrid24}.  Here we focus only on the detector noise that is composed of three components:
\begin{equation} \label{eq:det_noise}
\sigma_{detector}^2 = \sigma_{dark curret}^2 + \sigma_{depfet}^2 +\sigma_{asic}^2
\end{equation}
The dark current noise is the shot noise due to the thermal generation of charge carriers that occurs in the silicon bulk of the sensor. It is a crucial component because it is expected to experience the strongest change during the mission. It is equal to the square root of the number of generated electrons per readout cycle.
\begin{equation} \label{eq:sig_noise}
\sigma_{dark currentl}  = \omega\sqrt{g\cdot t} \, ,
\end{equation}
with frame time t, generation rate g and mean electron hole pair creation energy $\omega$. Here $\omega$ = 3.71 eV is used\cite{lowe07}.

With the measurement setup in this experiment the detector noise is measured and the three components of the detector noise can be determined individually. When the source to drain voltage of the DEPFET is set to 0~V the current flow from the sensor to the readout ASIC is switched off. The remaining noise measured with the readout system is approximately equal to $\sigma_{asic}$ (with some additional minor contributions of the ADCs and a buffer between the readout ASIC and the ADCs).
The dark current noise $\sigma_{dark current}$ is measured by varying the frame time in the range from 0.25 to 9 ms and fitting the measured total noise according to equation \ref{eq:sig_noise}. The DEPFET noise is then obtained by subtracting the dark current noise and asic noise from the total detector noise according to equation \ref{eq:det_noise}. 

   \begin{figure} [ht]
   \begin{center}
   \begin{tabular}{cc}
   \includegraphics[height=0.27\textwidth]{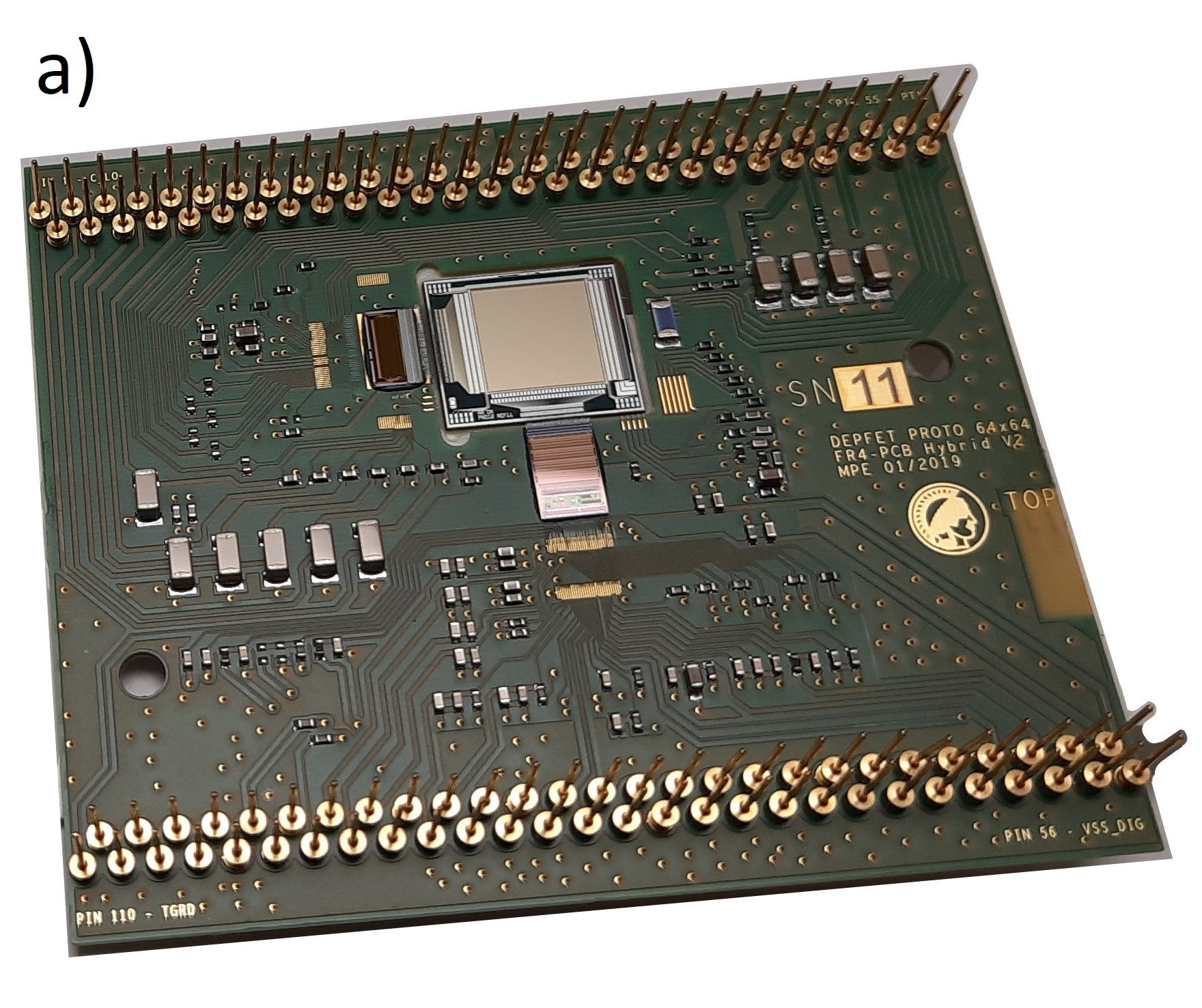}
   \includegraphics[height=0.27\textwidth]{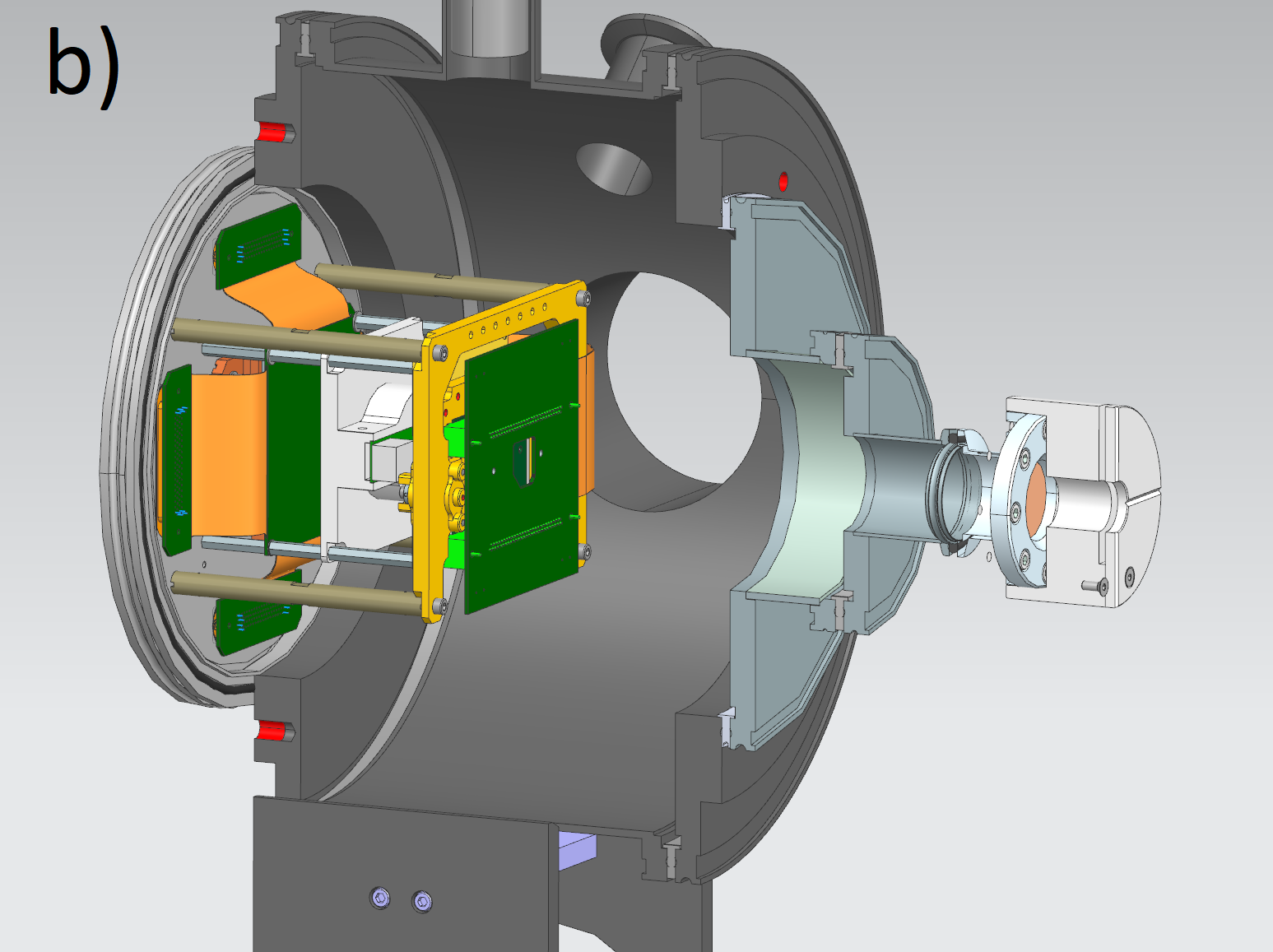}
	\end{tabular}
	\end{center}
   \caption[] 
   { \label{fig:chamber}
 Left: Picture of a detector module like the ones that were used for the irradiation tests. Right: Setup with vacuum chamber that was used.}
   \end{figure}

In orbit the dark current noise increases due to lattice defects that are introduced through particle radiation, i.e. the Total Non Ionizing Dose (TNID). The TNID that results from a particle depends on the particle type and energy. In the scope of the NIEL scaling theory they can be compared to each other under the assumption that the damage to the lattice is proportional to the Non-Ionizing Energy Loss (NIEL) of the particle. It is then possible to express a certain TNID as a displacement damage equivalent fluence (DDEF) that is defined as follows:
\begin{equation} \label{eq:ddef}
\Phi_{eq} = \kappa \: \Phi_{p,E} \, , 
\end{equation}
where $\Phi_{eq}$ is the equivalent flux, $\kappa$ the so called hardness factor and $\Phi_{p,E}$ the flux of particles of type p and energy E. The equivalent flux is usually defined for neutrons with an energy of 1 MeV. In this work the equivalent flux of 10-MeV protons is used.
The hardness factor $\kappa$ is given by:
\begin{equation} \label{eq:hardness}
\kappa = \frac{NIEL(E)_{p}}{NIEL(E_{eq})_{p_{eq}}} \, ,
\end{equation}
where $NIEL(E)_{p}$ is the non-ionizing energy loss of a particle of type p and energy E, $E_{eq}$ and $p_{eq}$ are the energy and particle type for which the equivalent flux is defined\cite{mixs_rate}.

As the dark current $I_{dark}$ results from thermal excitation of charge carriers it shows a very strong temperature dependence. The ratio of the dark current at temperatures ${T_1}$ and ${T_2}$ is usually given by 
\begin{equation} 
\label{eq:Tdep}
\frac{I_{dark}(T_1)}{I_{dark}(T_2)} = \left( \frac{T_1}{T_2} \right)^2\:e^{\frac{E_g}{2k_B} \left( \frac{1}{T_1}-\frac{1}{T_2} \right) } \, ,
\end{equation}
 where $k_B$ is the Boltzmann constant\cite{Sze81}. 
At room temperature the parameter E\textsubscript{g} is usually assumed to be identical to the band gap energy of 1.12 eV, but there are cases where lower values are found to produce a better agreement with the measurement data. A possible explanation for this is given in Ref.~\citenum{darkCCD}. The authors argue that at low temperatures changing contributions from depletion dark current and diffusion dark current can lead to a temperature dependence of E\textsubscript{g}.

\section{Proton irradiation}
\label{sec:proton}  % \label{} allows reference to this section

\subsection{Experiment}
\label{sec}

The irradiation took place at the synchrotron facility MedAustron in Wiener Neustadt, Austria. Two detector modules were irradiated, each comprising a 64$\times$64 pixel DEPFET sensor glued to a ceramic carrier. The carrier is attached to a PCB that holds a readout ASIC (VERITAS 2.2~\cite{sven18, anna24}), a steering ASIC (Switcher A~\cite{fischer03}), and several passive components (Figure~\ref{fig:chamber}a). The DEPFET sensors originated from a wafer of the first NewAthena flight-production batch. \

During irradiation, the detectors were mounted in a vacuum chamber (see Figure~\ref{fig:chamber}b). The chamber is equipped with a Stirling cooler that cools the entire detector module down to the operating temperature specified for the sensor. Power supply, steering, and readout are provided through feedthroughs on the rear side of the chamber. These are connected via a flexible lead to an adapter board that holds the detector module. The adapter board has an opening large enough to irradiate the front side of the detector and the ASICs without any additional material in the beam path. With this setup, the detectors can be fully biased and operated during irradiation. The protons pass through the exit window of the beamline (equivalent to 2.4 mm of water) and pass through 85~cm of air before they enter the chamber through a dedicated entrance window that must be both very thin and light-tight. It consists of a 50~µm polyimide foil with a nickel coating and therefore causes only a minimal proton energy loss (\textless 70~keV at 7~MeV).\

The first detector was irradiated with 62.4~MeV protons using the highest available beam flux. In this operating mode, which is routinely employed for medical applications, the accelerator provides a calibrated online dosimetry system. Ten proton spills of 1~s duration, separated by 1~s pauses, were delivered, corresponding to a total of 5.94$\cdot$10\textsuperscript{10} protons according to the accelerator dosimetry. Since the detector area is smaller than the beam cross section, the absolute dose received by the sensor cannot be determined accurately from these data alone. This irradiation primarily served to investigate possible transient detector effects at extremely high dose rates and to study the short-term annealing of radiation-induced defects at the nominal operating temperature of 213~K. \

A second detector was irradiated with 62.4~MeV protons using the highest of the three available low-flux settings~\cite{pur21}. This irradiation was performed similarly to that of the pre-flight production presented in Ref.~\citenum{vxe24}; additional information on the accelerator facility, the dosimetry method, and the measurement setup can be found there. The primary objective of this irradiation was to establish the relationship between TNID and radiation-induced dark current.\

   \begin{figure} [ht]
   \begin{center}
   \begin{tabular}{c} 
   \includegraphics[width=0.90\textwidth]{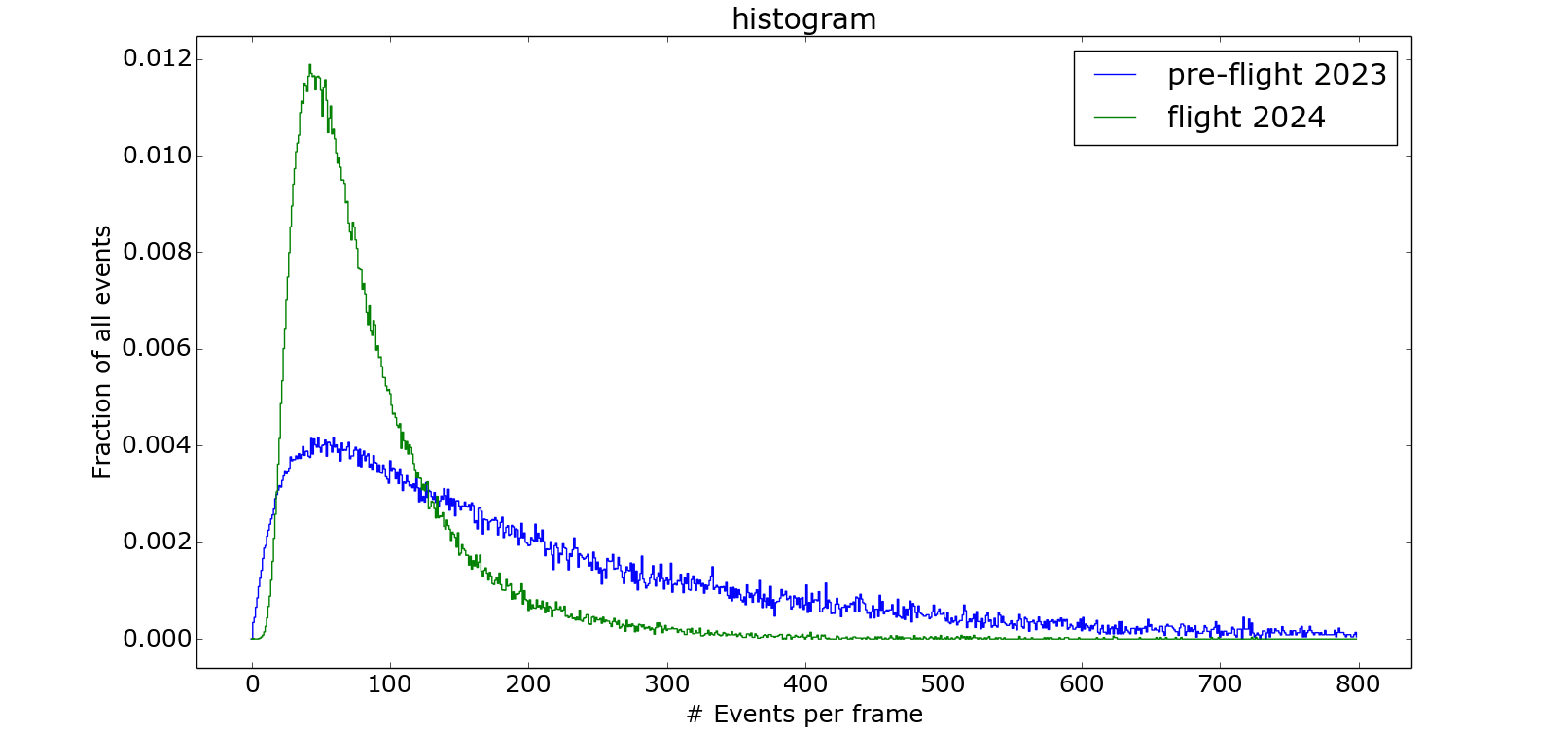}
	\end{tabular}
	\end{center}
   \caption[] 
   { \label{fig:Intensity_hist} 
The distribution of the number of events per frame illustrates the extent of the flux variations.}
   \end{figure}

The goal was to keep the proton flux low enough to enable dosimetry using the DEPFET sensor itself, while still being high enough to reach at least the expected dose during the goal lifetime of 10 years. The synchrotron delivered proton spills of 10~s duration followed by 3~s pauses, for a total beam-on time of 5.5~h. Although the average proton flux was well defined, the instantaneous flux within individual spills exhibited fluctuations of unspecified amplitude and frequency. Consequently, only mean flux values are reported.\

Previous irradiation studies of pre-flight production sensors, as well as the high-flux experiment described above, had already demonstrated that radiation-induced dark current continues to decrease at 213~K owing to the annealing of lattice defects. To investigate this behaviour quantitatively, the detector was maintained at 213~K for 20 days after irradiation while dark frame measurements for the determination of the noise were done at regular intervals. This was possible by using an uninterruptible power while the setup was transferred to the adjacent decay room. Since this room is only accessible intermittently and does not provide a network connection, the data acquisition system was configured for fully autonomous operation. At the end of the 20-day cold annealing period, the detector noise was characterized over the relevant operating temperature range.

\subsection{Dose and dose rate}
\label{sec:dose1}

The dosimetry for the second detector was based entirely on the radiation detected by the device under test. The number of protons incident on each pixel can be determined from the recorded data using two different approaches. The first approach applies the standard event filtering and recombination algorithms described in Refs.~\citenum{robert08, lauf14}. This method is applicable only when the proton flux is sufficiently low that event pileup can be neglected. While this condition is not completely met for the average proton flux over the complete irradiation (approximately 3.8$\cdot$10\textsuperscript{5}~62.4-MeV~protons/cm\textsuperscript{2}$\cdot$s, corresponding to 42.4~protons/frame), short-term beam intensity fluctuations are causing a fraction of the frames to be severely affected by pileup. 

   \begin{figure} [ht]
   \begin{center}
   \begin{tabular}{c} 
   \includegraphics[width=0.8\textwidth]{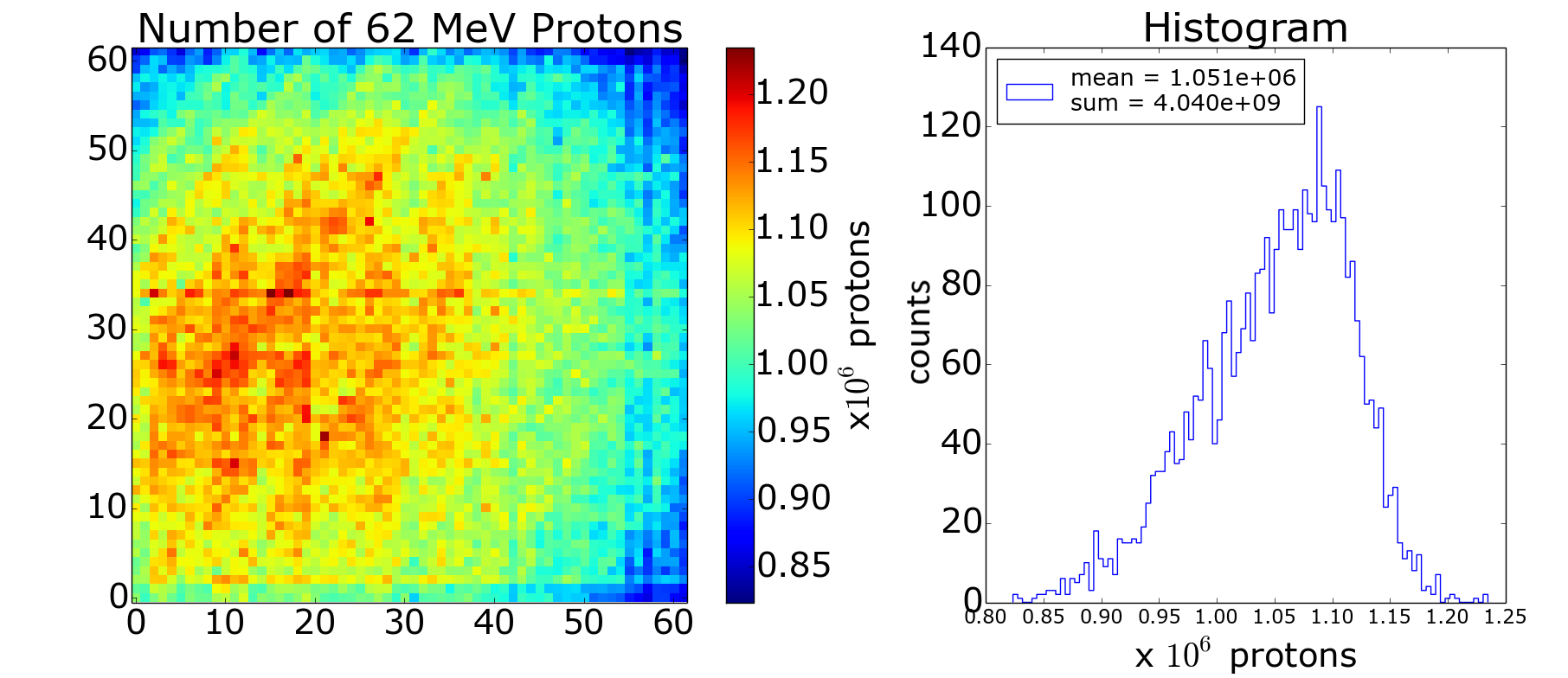}
	\end{tabular}
	\end{center}
   \caption[Number of protons] 
   { \label{fig:proton_map} 
Left: Map of the number of protons per pixel. The FWHM of the beam profile is clearly larger than the sensor. Right: The histogram shows the variation of the total fluence over the sensor area.}
   \end{figure}

The settings of the low-flux modes of the accelerator were modified compared to the previous radiation tests with the pre-flight production. This resulted in smaller short-term flux variations. However, the distribution shown in Figure~\ref{fig:Intensity_hist} demonstrates that there is still a significant number of events that originate from frames that are significantly affected by pileup.
 
 For this reason, the proton fluence is determined from the integrated deposited energy rather than by directly counting individual proton events. Following the method described in Ref.~\citenum{vxe24}, the mean signal per event is first determined and subsequently used to convert the integrated signal recorded in each pixel into the corresponding number of incident protons. The resulting mean signal amounts to 778~\textpm~1~keV/event. For frames with high occupancy, this value decreases by approximately 98~eV per event owing to event overlap.

The resulting proton map is shown in Figure~\ref{fig:proton_map}. The mean proton fluence corresponds to 1.05$\cdot$10\textsuperscript{6} protons per pixel. The diagonal stripe pattern originates from bulk doping variations within the sensor material, which cause small variations in the effective pixel area. The same pattern is observed in flat-field X-ray measurements, confirming its detector origin. Since the associated intensity modulation is only of the order of a few percent, it has a negligible influence on the derived proton fluence.

For comparison, direct proton counting yields approximately 7\% fewer protons because overlapping charge patterns are reconstructed as single events in frames with high occupancy.

The measured proton fluence is converted into a DDEF of 10-MeV protons using Equation~\ref{eq:ddef} together with the hardness factor defined in Equation~\ref{eq:hardness}. As demonstrated in Ref.~\citenum{vxe24}, the hardness factor for monoenergetic 62.4~MeV protons ($\kappa_\text{eff}=0.417~\pm~0.003$) provides an accurate description for the present irradiation conditions.

The mean displacement damage dose corresponds to 2.6$\cdot$10\textsuperscript{9}~10-MeV~protons/cm\textsuperscript{2}. Considering the effective beam-on time of 15230~s (excluding the 3~s pauses between the 10~s spills), the corresponding mean dose rate is 1.7$\cdot$10\textsuperscript{5}~10-MeV~protons/cm\textsuperscript{2}$\cdot$s.

\subsection{Results}
\label{sec:results}

\begin{figure} [ht]
   \begin{center}
   \begin{tabular}{c} 
   \includegraphics[width=0.75\textwidth]{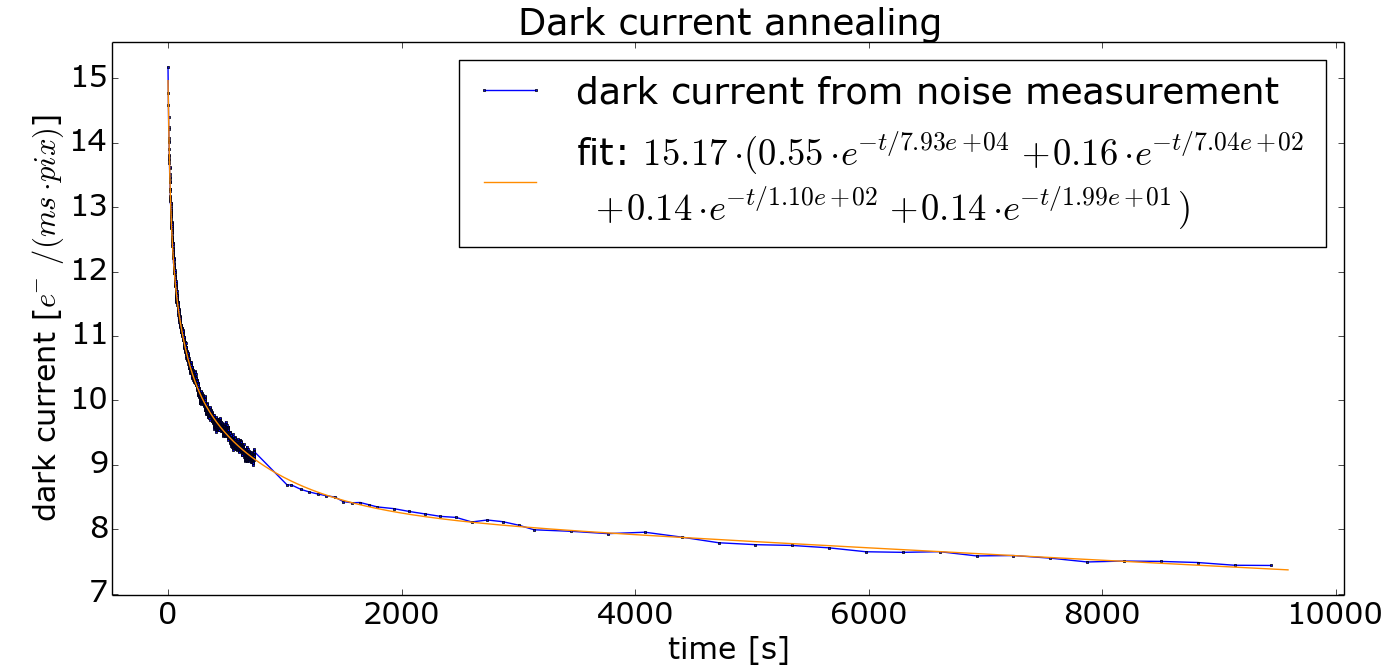}
	\end{tabular}
	\end{center}
   \caption[dark current annealing] 
   { \label{fig:annealing1} 
Dark current annealing of a detector module after short irradiation (19 s) with a very high dose rate. At least 4 different time constants were necessary to fit the evolution of the dark current during the first 9500 s after the irradiation.}
   \end{figure}

\begin{figure} [ht]
   \begin{center}
   \begin{tabular}{c} 
   \includegraphics[width=0.75\textwidth]{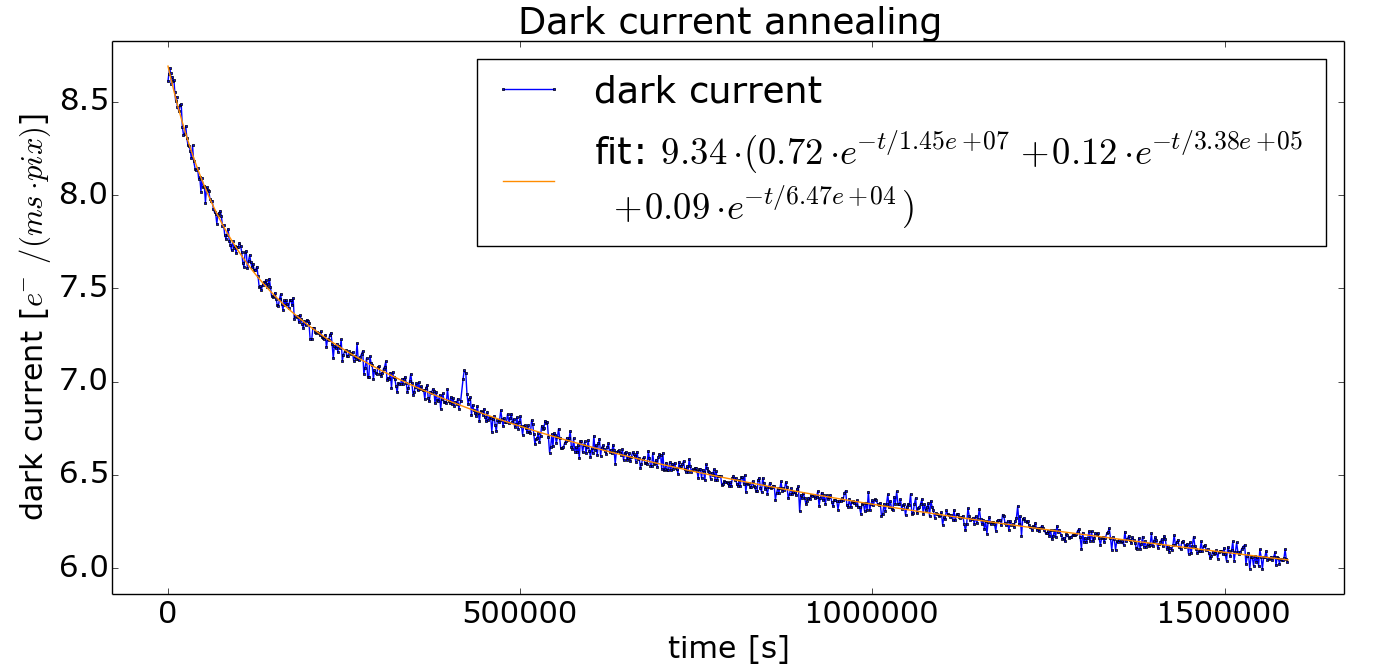}
	\end{tabular}
	\end{center}
   \caption[dark current annealing] 
   { \label{fig:annealing2} 
Dark current annealing of the detector module that was irradiated with low dose rate. At least 3 different time constants were necessary to fit the evolution of the dark current during the first 18.4 days after the irradiation.}
   \end{figure}

Immediately after irradiation, the total noise of both detector modules was monitored over time. Assuming that the radiation exposure affects only the dark current noise while all other noise components remain unchanged, the radiation-induced dark current can be estimated by subtracting the pre-irradiation noise from the measured total noise. Although this approach does not provide the most accurate absolute values of the dark current, it allows its temporal evolution to be followed with excellent time resolution. Figures~\ref{fig:annealing1} and~\ref{fig:annealing2} show the evolution of the radiation-induced dark current after irradiation for both detectors. In one case, the first 9500~s are covered, while for the other detector the measurement extends over 20~days.\

   
 The microscopic processes responsible for the observed annealing time constants are of considerable scientific interest but are beyond the scope of this work. For the present study, it is sufficient to conclude that annealing of radiation-induced lattice defects causes a substantial reduction of the dark current even at the nominal operating temperature of 213~K. The measured time constants span almost six orders of magnitude, from approximately 20~s to 1.5$\cdot$10\textsuperscript{7}~s. Longer time constants cannot be resolved within the available observation period. Consequently, accurately predicting the equilibrium dark current under the continuous low-dose-rate irradiation expected in orbit is not straightforward. Nevertheless, performing the final dark current characterization after 20~days of annealing provides a conservative estimate of the detector dark current expected after 10~years of operation in space.\
 
\begin{figure} [ht]
   \begin{center}
   \begin{tabular}{cc}
   \includegraphics[width=0.46\textwidth]{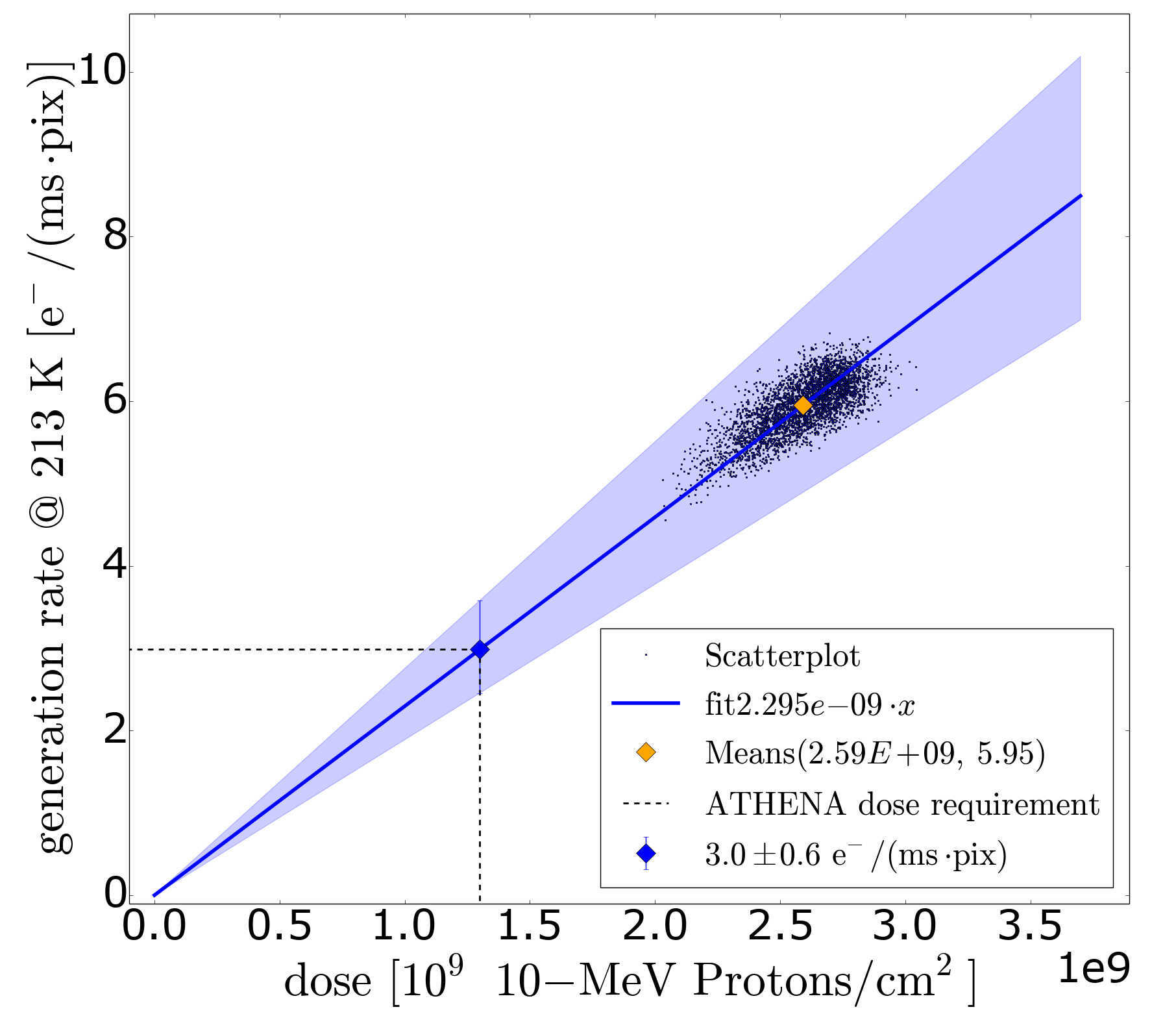}
   \includegraphics[width=0.53\textwidth]{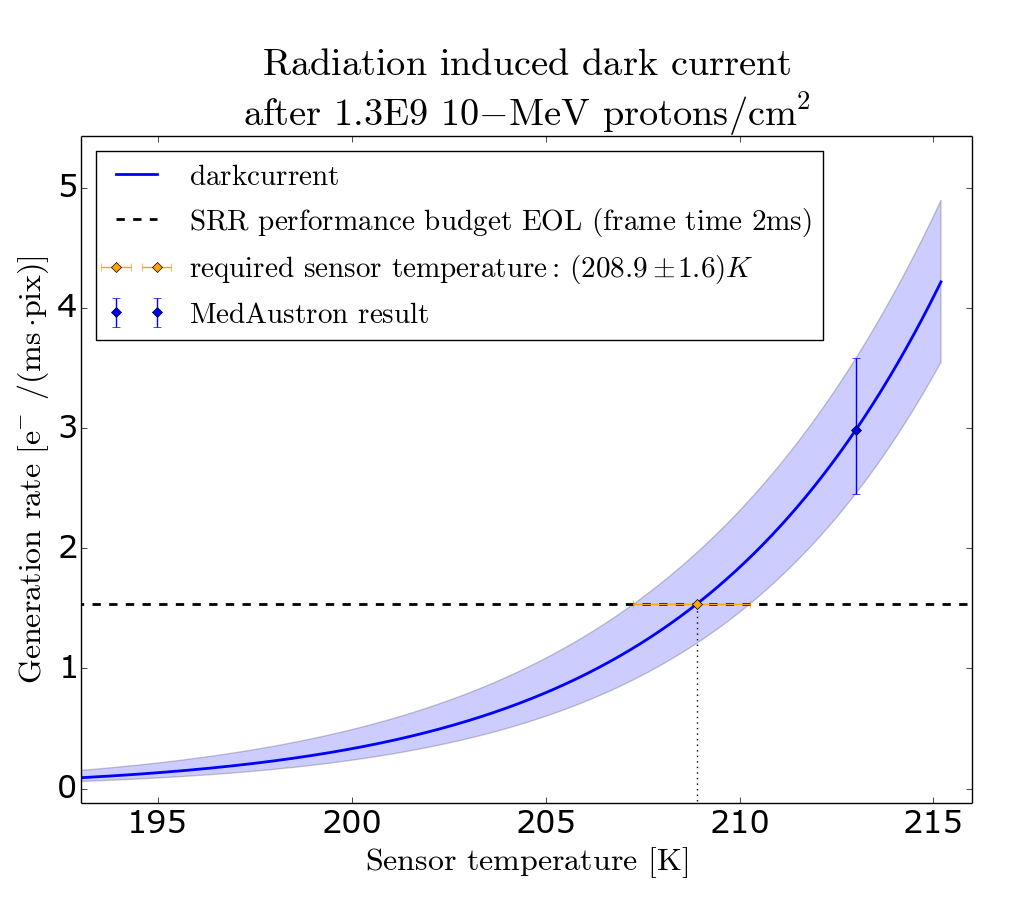}
   \end{tabular}
   \end{center}
   \caption[] 
   { \label{fig:dose_curr}
 Left: Scatterplot of the dose and dark current values of all pixels. The blue dot represents the expected dark current at the expected dose after 10 years in space and thus the main result of this experiment. Right: The blue dot from the left figure is plotted with error bar that results from the uncertainty of the dose and current measurements. The dashed lines shows the dark current budget for a frame time of 2~ms, the dotted line the necessary sensor temperature for that value.}
   \end{figure}

The left panel of Figure~\ref{fig:dose_curr} shows the correlation between radiation-induced dark current and displacement damage dose for all pixels. A linear fit to the complete data set yields the proportionality between dose and radiation-induced dark current. The shaded region represents the uncertainty of the fitted slope, taking into account the uncertainties of both the dose and dark current measurements. Extrapolation to the expected end-of-life dose after 10~years in orbit predicts a radiation-induced dark current of 3.0~\textpm~0.6 electrons per pixel and millisecond. Combining this result with the measured temperature dependence of the dark current allows the required operating temperature at EOL to be estimated. As shown in the right panel of Figure~\ref{fig:dose_curr}, the detector temperature must remain below (208.9~\textpm~1.6)~K for operation at a frame rate of 500~Hz.
 
\begin{figure} [ht]
   \begin{center}
   \begin{tabular}{cc} 
   \includegraphics[width=0.48\textwidth]{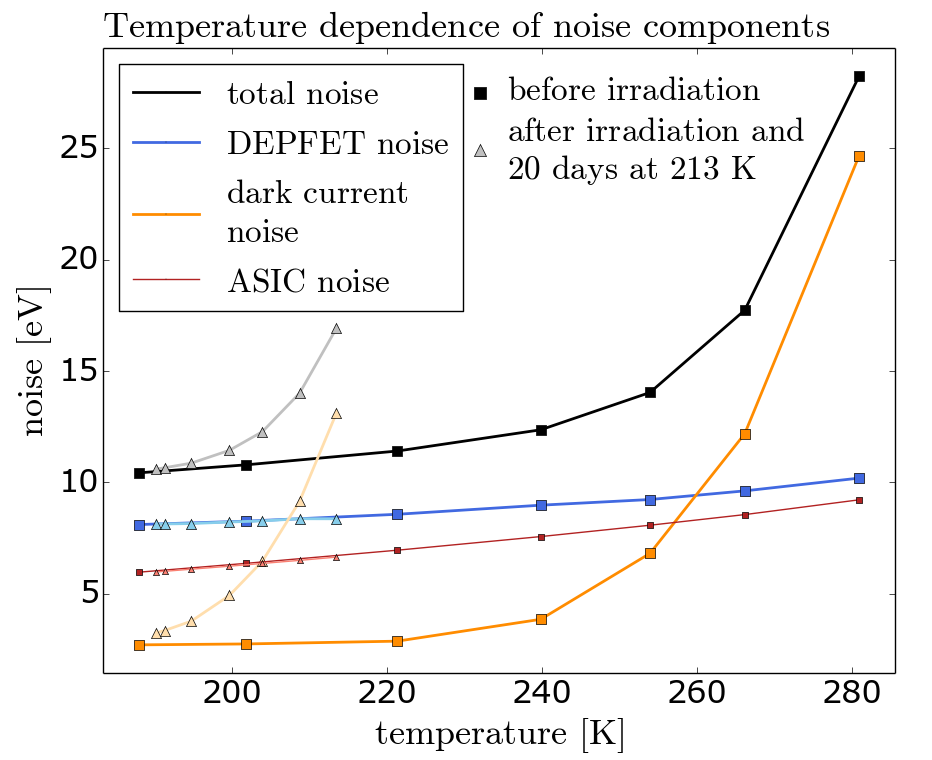}
   \includegraphics[width=0.48\textwidth]{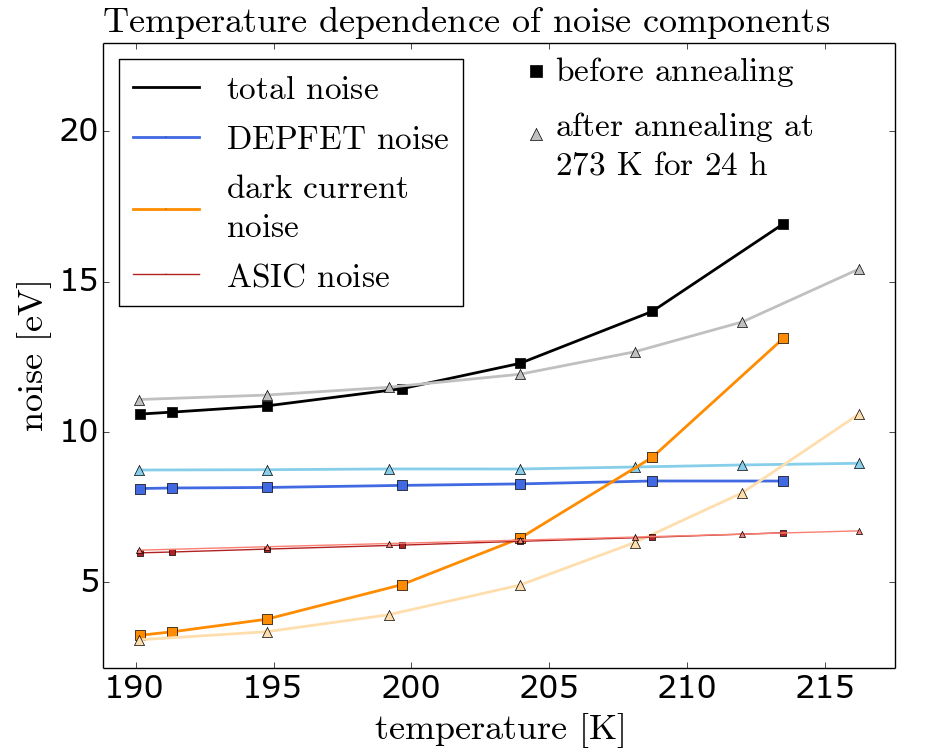}
	\end{tabular}
	\end{center}
   \caption[] 
   { \label{fig:noise_comp} 
Results of the temperature dependent noise characterization. In the left panel the pre-irradiation measurements are compared to post-irradiation measurements that were done after the detector stayed at 213~K for 20~days. The second characterization was only done for temperatures at or below 213~K in order to avoid any unwanted annealing that could influence the results. In the right panel measurements before and after annealing at 273~K for 24~h are compared. }
   \end{figure}

A temperature-dependent characterization of the detector noise was carried out before irradiation and repeated after 20~days of annealing at 213~K. The comparison is shown in Figure~\ref{fig:noise_comp}. As expected, proton irradiation increases only the dark current noise associated with thermally generated dark current, while the remaining noise components are initially unaffected. After annealing at 273~K for 24~h, however, the dark current noise decreases whereas the intrinsic DEPFET noise increases measurably. As discussed in Section~\ref{sec:TID}, this additional DEPFET noise originates from the total ionizing dose deposited by the protons in the gate oxide. At temperatures below approximately 200~K, the increase in DEPFET noise exceeds the reduction in dark current noise, resulting in an overall increase of the total detector noise.

\section{TID Test}
\label{sec:TID}  % \label{} allows reference to this section
\subsection{Experiment}
\label{sec:tid_setup}  % \label{} allows reference to this section

The total ionizing dose (TID) irradiation campaign closely followed the procedure previously established for the pre-flight production detectors~\cite{vale22}. Additional details are given in Ref.~\citenum{luisa25}. Another detector module equipped with a 64$\times$64 pixel DEPFET sensor from an NewAthena flight-production wafer was used.

The irradiation was performed using a standard analytical X-ray tube mounted in a Seifert SH 37/80 RÖ housing and powered by a GE/Seifert ISO-DEBYEFLEX 3003 generator. The X-ray source was attached to the same vacuum chamber that had been used for the proton irradiation experiments. The Mo target was preferred over Fe and Cu targets in order to obtain higher-energy characteristic X-rays. This results in a more uniform depth profile of the energy deposition from the absorbed radiation in the device under test. The tube was operated at an acceleration voltage of 30~kV, while the dose rate was adjusted by varying the anode current between 2 and 50~mA. A filter wheel positioned between the X-ray tube and the detector allowed different filters to be inserted into the beam. For the irradiation, relatively thick ($\sim$100~µm) zirconium and molybdenum filters were used, producing a spectrum dominated by the Mo-K\textsubscript{$\alpha$} emission line.

The primary objective of this irradiation campaign was to identify the physical origin of the additional noise component that had been observed after room-temperature annealing in the previous TID study. A second objective was to investigate both the gradual development of this noise contribution after irradiation and its dependence on temperature. In contrast to the previous experiment, the readout ASICs were shielded from the X-rays by a 2~mm thick steel plate so that only the DEPFET sensor was irradiated.

Instead of determining the threshold voltage shift for every individual pixel, only the mean threshold voltage shift of the detector was monitored. This was accomplished by supplying the source contact of the DEPFET array to a Source Meter Unit (SMU), while the drain potential was fixed by the readout ASIC. The SMU operated in constant-current mode, so that a radiation-induced shift of the threshold voltage resulted directly in a corresponding change of the source voltage. As long as the transistor remained close to its nominal operating point, the measured source-voltage shift provided a good approximation of the mean threshold voltage shift of the DEPFET array.

\subsection{Dose and dose rate}
\label{sec:dose2}

The interaction of the WFI instrument with the expected radiation environment at L2 has been investigated in a series of dedicated simulation studies~\cite{avk18, tanja20, tanja21}. These simulations were primarily performed to optimize the instrument design with respect to the non-X-ray background. Using the corresponding detector mass model and simulation framework, the expected total ionizing dose (TID) was also evaluated. For a ten-year mission starting in 2028, a TID of 4.97~\textpm~0.10~Gy(Si) was predicted at a confidence level of 95\%. Although the mission baseline has since changed from an L2 orbit to L1 and the launch date is currently scheduled for 2039, the expected TID is sufficiently similar that the original prediction remains an appropriate reference for the present irradiation campaign.\

\begin{figure} [ht]
   \begin{center}
   \begin{tabular}{c} 
   \includegraphics[width=0.8\textwidth]{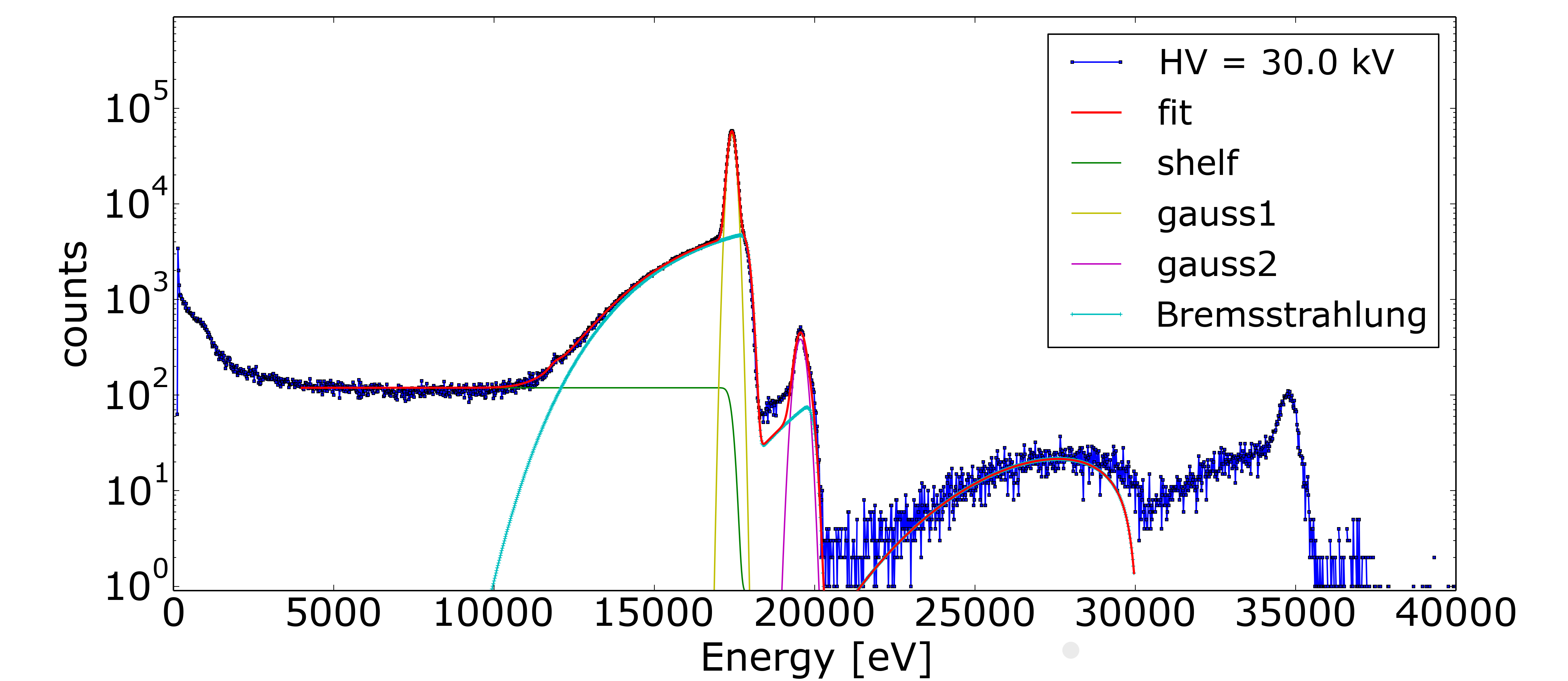}
	\end{tabular}
	\end{center}
   \caption[meas spec] 
   { \label{fig:meas_spectrum} 
Measured X-ray spectrum and fit.}
   \end{figure}

\begin{figure} [ht]
   \begin{center}
   \begin{tabular}{c} 
   \includegraphics[width=0.8\textwidth]{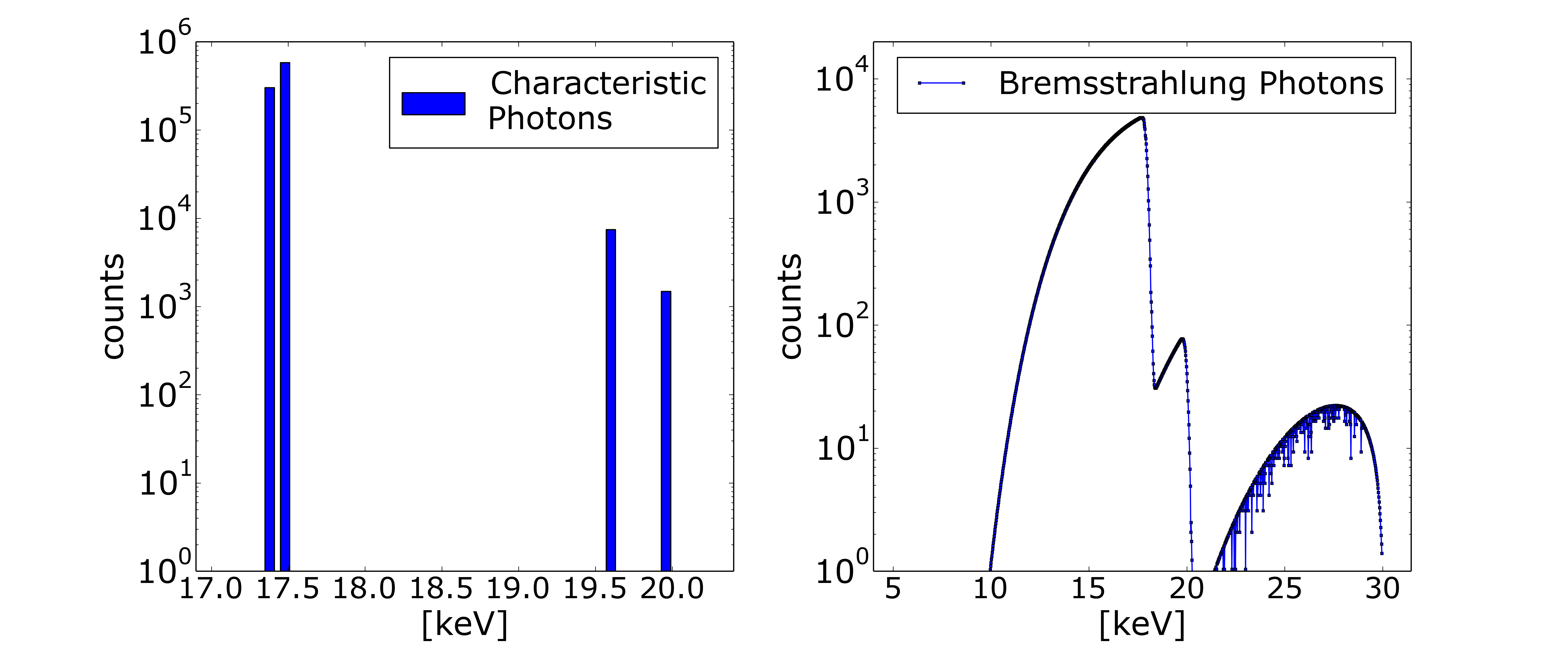}
	\end{tabular}
	\end{center}
   \caption[in spec] 
   { \label{fig:input_spectrum} 
The reconstructed incident X-ray spectrum. }
   \end{figure}

The dosimetry for the TID irradiation was performed using the DEPFET detector itself. The procedure used to derive the absorbed dose from the detector signal is described in detail in Ref.~\citenum{vale22}. The conversion between deposited signal and absorbed dose depends on the spectral distribution of the incident X-rays but is independent of the photon flux and therefore of the selected anode current. Figure~\ref{fig:meas_spectrum} shows the spectrum measured with the detector, while Figure~\ref{fig:input_spectrum} presents the incident X-ray spectrum reconstructed using the detector-response model. The resulting dose-to-signal conversion factor is 13.09~mGy/(GeV/pix).

\begin{figure} [ht]
   \begin{center}
   \begin{tabular}{cc} 
   \includegraphics[width=0.48\textwidth]{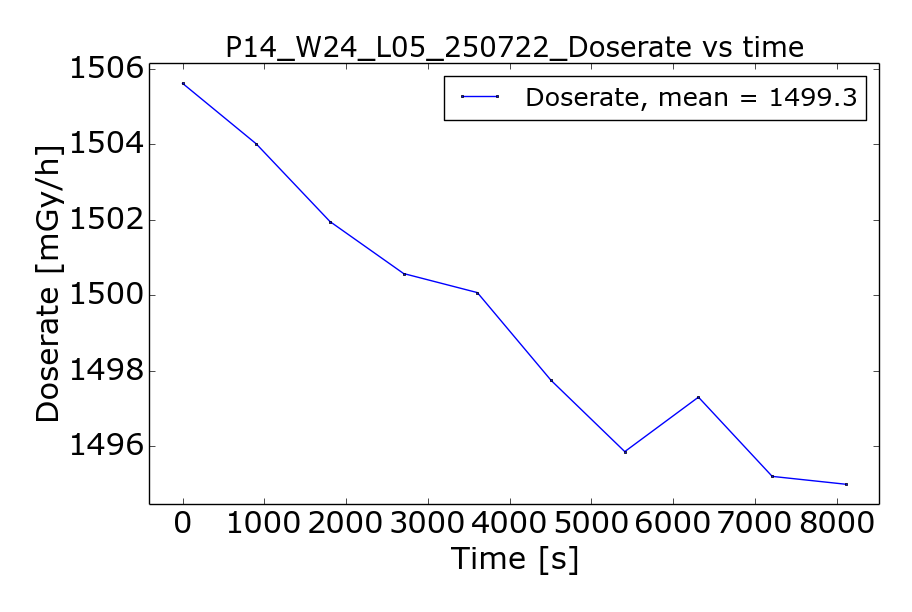}
   \includegraphics[width=0.48\textwidth]{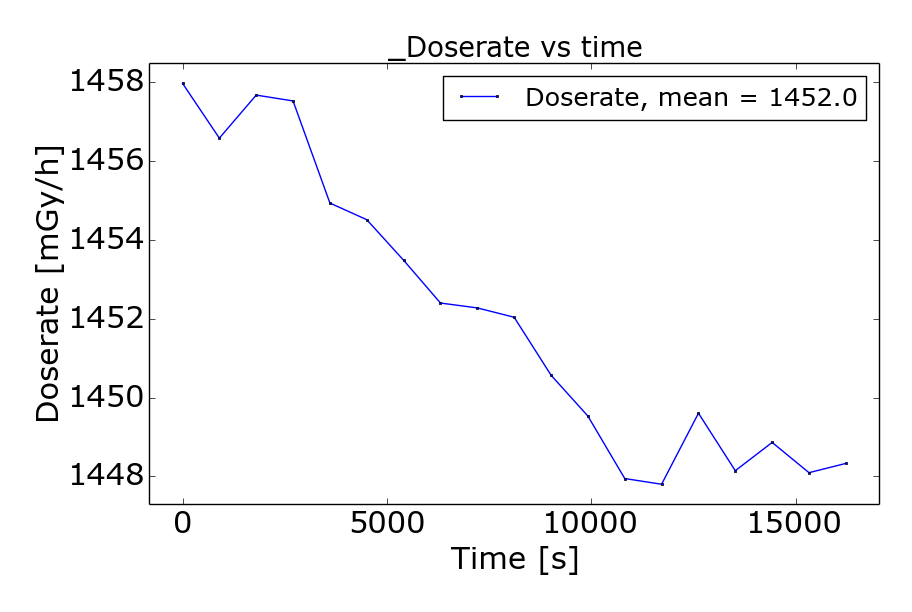}
	\end{tabular}
	\end{center}
   \caption[] 
   { \label{fig:dose_rate} 
 Dose rates during the irradiations. For the first irradiation dose rates were only measured during the first half. }
   \end{figure}

The irradiation was carried out in two 5~h sessions on consecutive days. During each irradiation, the instantaneous dose rate was monitored by recording 3000 detector frames at regular intervals (Figure~\ref{fig:dose_rate}). Integrating the measured dose rate over both irradiation periods yields a total delivered dose of 15~Gy.

\subsection{Results}
\label{sec:results2}

\begin{figure} [ht]
   \begin{center}
   \begin{tabular}{c} 
   \includegraphics[width=0.9\textwidth]{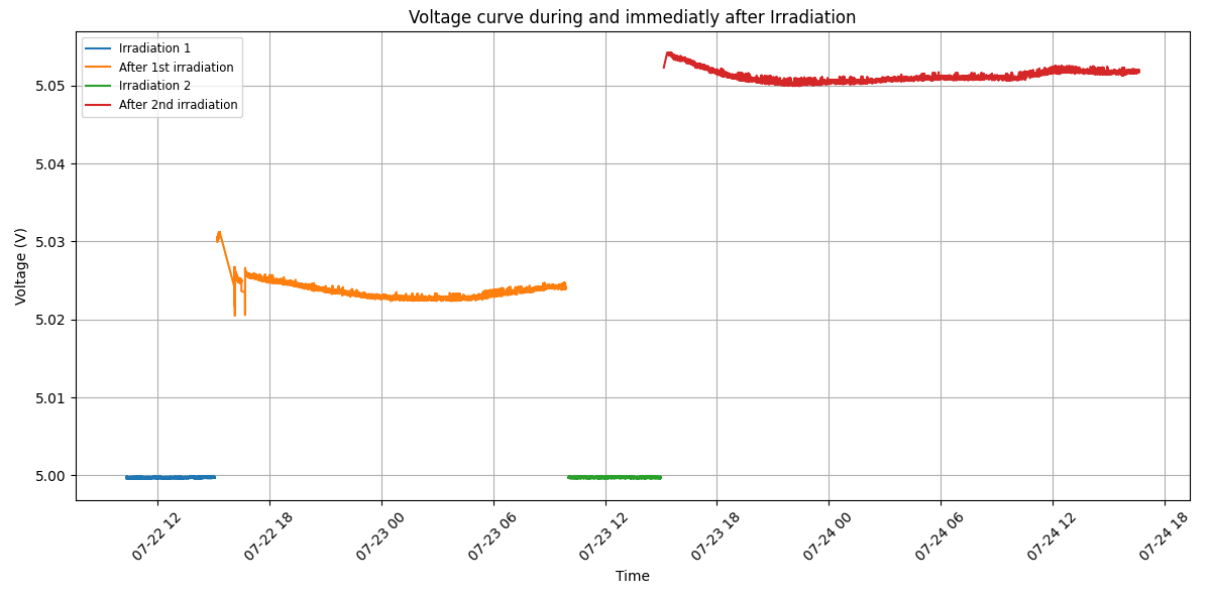}
	\end{tabular}
	\end{center}
   \caption[threshold voltage] 
   { \label{fig:thres_volt}   
DEPFET source voltage measurement that was used to monitor the threshold voltage shift. Prior to irradiation, the gate voltage was adjusted such that V\textsubscript{Source} = 5.00~V at a channel current of 100~µA per pixel. The SMU was operated in constant-current mode, maintaining I\textsubscript{Source-Drain} = 100~µA per channel. Under these conditions, the source voltage directly tracks changes in the threshold voltage. During irradiation, the source voltage was fixed at 5~V. Figure taken from \citenum{luisa25}.}
   \end{figure}

The evolution of the threshold voltage is shown in Figure~\ref{fig:thres_volt}. After the full irradiation dose, the mean threshold voltage exhibits a shift of 50~mV. The detector performance is unaffected immediately after irradiation. Table~\ref{tab:TID} summarizes the results of a series of dark frame and \textsuperscript{55}Fe measurements. As in the proton irradiation campaign, the device was kept at the nominal operating temperature of 213~K after irradiation. From Table~\ref{tab:TID}, a gradual increase in readout noise is already observable during the first three weeks following irradiation.

\begin{table}[ht]
\caption{Results of dark frames and \textsuperscript{55}Fe measurements before and after irradiation and annealing.} 
\label{tab:TID}  
\begin{center}       
\begin{tabular}{|l|l|l|l|l|} %% this creates five columns
%% |l|l| to left justify each column entry
%% |c|c| to center each column entry
%% use of \rule[]{}{} below opens up each row
\hline
\rule[-1ex]{0pt}{3.5ex}   & Mean Noise  & FWHM @ 5.9 keV  & Gain        & Source Voltage\\
\rule[-1ex]{0pt}{3.5ex}   & [$e^-$ ENC] &  [eV]                       & [adu/eV] & [V]\\
\hline
\rule[-1ex]{0pt}{3.5ex} Before irradiation        & 2.891 & 129.4 & 0.837 & 5.000\\
\hline
\rule[-1ex]{0pt}{3.5ex} After 1st irradiation     & 2.896 & 129.4 & 0.836 & 5.025\\
\hline
\rule[-1ex]{0pt}{3.5ex} After 2nd irradiation   & 2.902 & 129.6 & 0.838 & 5.050\\
\hline
\rule[-1ex]{0pt}{3.5ex} Three weeks later        & 3.098 & 131.2 & 0.833 & 5.050\\
\hline
\rule[-1ex]{0pt}{3.5ex} After 1 week @ 273 K & 3.245 & 131.8  & 0.835 & 5.050\\
\hline
\end{tabular}
\end{center}
\end{table}

   \begin{figure} [ht]
   \begin{center}
   \begin{tabular}{c} 
   \includegraphics[width=0.6\textwidth]{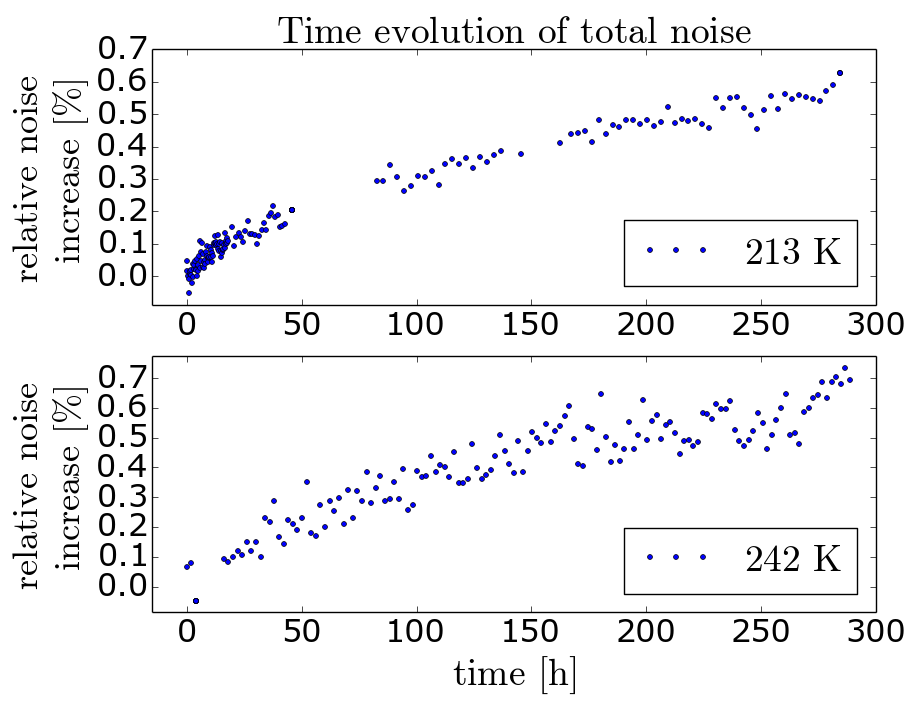}
	\end{tabular}
	\end{center}
   \caption[time evolution] 
   { \label{fig:anneal_TID} 
Top: The detector stayed at the operating temperature after irradiation. Noise measurements were done during the first 290 hours.
Bottom: After 22 days the temperature was raised to 242~K for 290 hours. Noise values are normalized to the mean of the first 4 measurements in both graphs.}
   \end{figure} 

The temporal evolution of the detector response is summarized in Figure~\ref{fig:anneal_TID}. During the initial 290~h at 213~K, the noise exhibits a progressive increase that gradually approaches saturation. After three weeks at 213~K, increasing the temperature to 242~K induces a further increase in noise of comparable magnitude over a similar time scale (Figure~\ref{fig:anneal_TID}, lower panel). This behavior is consistent with a thermally activated diffusion process governing the evolution of radiation-induced defects.

\begin{figure} [ht]
   \begin{center}
   \begin{tabular}{c} 
   \includegraphics[width=0.56\textwidth]{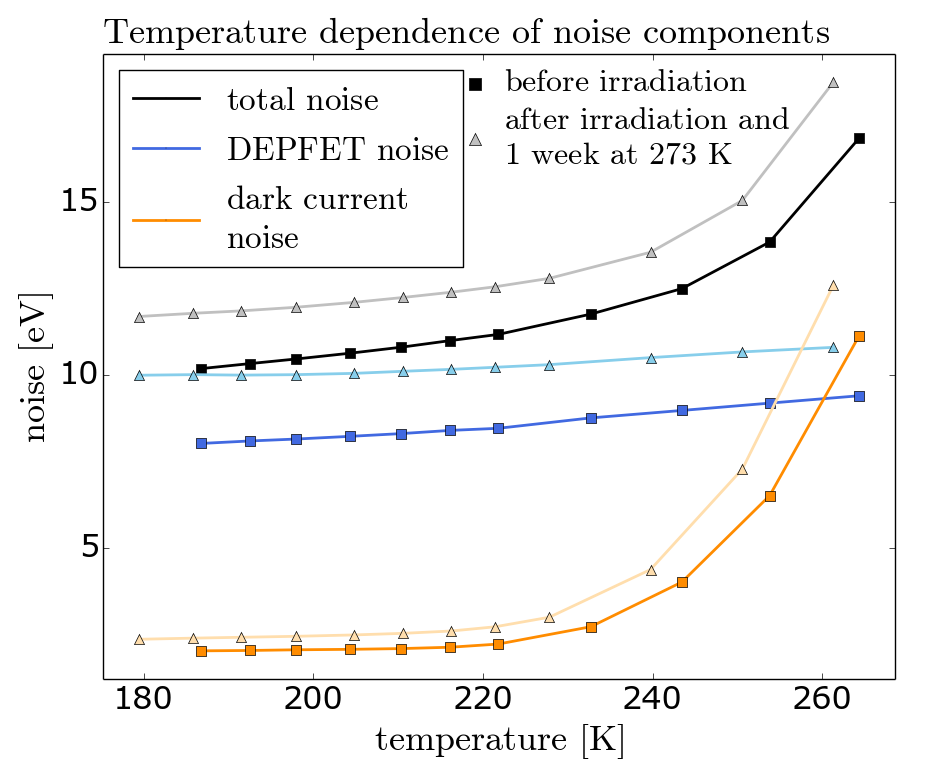}
	\end{tabular}
	\end{center}
   \caption[Mean signal per event] 
   { \label{fig:noise_comp_TID} 
Results of the temperature dependent noise characterization before irradiation and after annealing at 273 K. The DEPFET noise 
is significantly increased. }
   \end{figure}

In a subsequent step, the detector was held at 273~K for one week. In this regime, the total noise is dominated by dark current contributions, and residual variations primarily reflect small temperature fluctuations. Comparison of the temperature-dependent noise measurements before irradiation and after annealing at 273~K nevertheless reveals a clear increase in the intrinsic DEPFET noise (Figure~\ref{fig:noise_comp_TID}). An additional contribution to the dark current noise is also observed, but it does not significantly affect the total noise in the relevant temperature range.

It is therefore concluded that the observed effects are consistent with radiation-induced changes in the gate oxide, likely involving the diffusion of trapped holes and/or ions. These radiation-induced species can form interface traps at the Si/SiO\textsubscript{2} boundary (see Ref.~\citenum{oldham2012}, Chapter 2.2.5). Depending on their lifetime and the readout timing, such traps can contribute to an increase in both white noise and 1/f noise components in the DEPFET~\cite{KirtonUren}.

\section{DISCUSSION and SUMMARY}
\label{sec:discuss}  % \label{} allows reference to this section

The irradiation campaigns performed with sensors from the NewAthena flight production confirm the results previously obtained for the pre-flight production. For the expected end-of-life displacement damage dose of 1.3$\cdot$10\textsuperscript{9}~10-MeV-protons/cm\textsuperscript{2}, the radiation-induced dark current is predicted to increase by 3.0~$\pm$~0.6~e\textsuperscript{-}/(pix$\cdot$ms). This value agrees well with the corresponding result of 2.5~$\pm$~0.4~e\textsuperscript{-}/(pix$\cdot$ms) measured for the pre-flight production, demonstrating that the radiation tolerance of the flight sensors is fully consistent with earlier detector generations.

A total ionizing dose of 15~Gy produced a mean threshold voltage shift of 50~mV, compared with 80~mV after 14~Gy for the pre-flight production. More importantly, annealing at temperatures close to room temperature resulted in an additional DEPFET noise contribution of 5.7~eV, in agreement with the previously measured value of 5.4~eV. The available data indicate that this additional noise component develops during thermal annealing rather than immediately after irradiation and is therefore expected to remain small as long as the detector temperature is maintained well below room temperature. Nevertheless, its long-term evolution under the thermal conditions expected in orbit cannot be predicted with sufficient confidence yet. Consequently, this contribution should be included conservatively in the detector performance budget.

The current mission baseline foresees detector operation at 213~K during the early phase of the mission in order to minimize contamination risks. As radiation damage accumulates, the detector temperature can gradually be reduced to 193~K to suppress the increasing dark current. The present measurements demonstrate that operation at 193~K is sufficient to reduce the radiation-induced dark current at EOL to a negligible contribution to the total detector noise. At the same time, the development of the TID-induced DEPFET noise is expected to be strongly suppressed at this temperature.

\begin{figure} [ht]
   \begin{center}
   \begin{tabular}{c} 
   \includegraphics[width=0.7\textwidth]{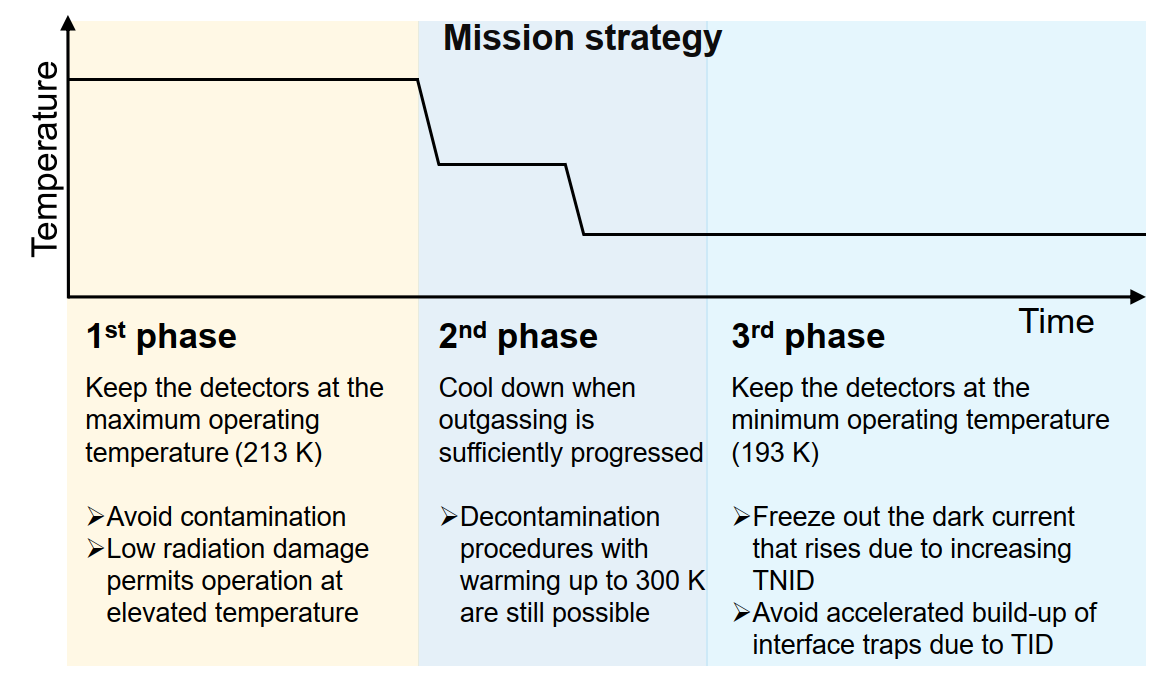}
	\end{tabular}
	\end{center}
   \caption[mission strategy] 
   { \label{fig:strategy}
 Optimal sensor temperature during the mission to minimize the impact of radiation damage on detector performance, under the assumption that the detector temperature must remain at 213~K during the early phase of the mission because of contamination concerns.}
   \end{figure}

This interpretation is supported by the previous TID study of the pre-flight production sensors, which was performed at 193~K and showed no measurable increase in DEPFET noise during the first two weeks after irradiation~\cite{vale22}. Consequently, provided that the detector temperature is reduced to 193~K sufficiently early in the mission, dedicated high-temperature annealing cycles during later mission phases appear neither necessary nor advantageous. While elevated temperatures would further reduce the dark current, this benefit is expected to be outweighed by the accompanying increase in intrinsic DEPFET noise. Based on the results presented in this work, the detector temperature profile illustrated in Figure~\ref{fig:strategy} is therefore recommended for operation throughout the NewAthena mission.

 \section*{ACKNOWLEDGMENTS} 
The authors thank Tomas Schreiner from MedAustron for access to and support at the accelerator facility. Development and production of the DEPFET sensors for the NewAthena WFI is performed in a collaboration between MPE and the MPG Semiconductor Laboratory (HLL). The work was funded by the German space agency DLR (FKZ: 50 QR 2301 and FKZ: 50 QR 2501).
 
% References
\bibliography{Eff_rad_damage_Athena_flight} % bibliography data in report.bib
\bibliographystyle{spiebib} % makes bibtex use spiebib.bst

\end{document}